\documentclass[prl,preprint]{revtex4}
\usepackage{epsfig}
\usepackage{graphicx}
\usepackage{dcolumn}
\usepackage{color}
\usepackage[T1]{fontenc}
\usepackage{amssymb, amsmath}
\usepackage{bm}

\definecolor{Gr}{rgb}{0,0.3,0}

\begin{document}
\bibliographystyle{apsrev}

\title{Long range coherent magnetic bound states in superconductors}

\author{Gerbold C. M\'enard$^1$}
\author{S\'ebastien Guissart$^2$}
\author{Christophe Brun$^1$}
\author{St\'ephane Pons$^{1,5}$}
\author{Vasily S. Stolyarov$^{1,3}$}
\author{Fran\c{c}ois Debontridder$^1$}
\author{Matthieu V. Leclerc$^1$}
\author{Etienne Janod$^4$}
\author{Laurent Cario$^4$}
\author{Dimitri Roditchev$^{1,5}$}
\author{Pascal Simon$^2$}
\email{pascal.simon@u-psud.fr}
\author{Tristan Cren$^1$}
\email{tristan.cren@upmc.fr}

\affiliation{$^1$Institut des Nanosciences de Paris, Sorbonne Universit\'es, UPMC Univ Paris 6 and CNRS-UMR 7588, F-75005 Paris, France}

\affiliation{$^2$Laboratoire de Physique des Solides, Universit\'e Paris-Sud, 91405 Orsay, France}

\affiliation{$^{3}$ Moscow Institute of Physics and Technology, 141700 Dolgoprudny, Russia}

\affiliation{$^4$Institut des Mat\'eriaux Jean Rouxel, CNRS Universit\'e de Nantes, UMR 6502, 2 rue de la Houssini\`ere, BP32229, 44322 Nantes, France}

\affiliation{$^5$Laboratoire de physique et d'\'etude des mat\'eriaux, LPEM-UMR8213/CNRS-ESPCI ParisTech-UPMC, 10 rue Vauquelin, 75005 Paris, France}

\date{\today}
%

\begin{abstract}

The quantum coupling of fully different degrees of freedom is a challenging path towards new functionalities for quantum electronics \cite{Thompson2013, Yeo2014,Yazdani}. Here we show that the localized classical spin of a magnetic atom immersed in a superconductor with a two-dimensional electronic band structure gives rise to a long range coherent magnetic quantum state. We experimentally evidence coherent bound states with spatially oscillating particle-hole asymmetry extending tens of nanometers from individual iron atoms embedded in a 2H-NbSe$_2$ crystal. We theoretically elucidate how reduced dimensionality enhances the spatial extent of these bound states and describe their energy and spatial structure. These spatially extended magnetic states could be used as building blocks for coupling coherently distant magnetic atoms in new topological superconducting phases \cite{Nadj-Perge2013,Choy2011,Nakosai2013,Braunecker2013, Klinovaja2013, Vazifeh2013,Pientka2013,Kim2013}.


\end{abstract}

\maketitle

Coupling different degrees of freedom of both quantum and classical objects yields new quantum functionalities not available in each system taken separately. In this regard, new hybrid quantum systems have been recently designed such as individual atoms and optical cavities coupled through photon exchange \cite{Thompson2013}, or a single quantum dot coupled to a mechanical oscillator \textit{via} strain \cite{Yeo2014}. A remarkable example of emerging phenomena is the observation of Majorana end states resulting from the coupling of a ferromagnetic chain of Fe atoms and a superconducting substrate \cite{Yazdani}. These states are in the focus of numerous theoretical works in the field of quantum computing as they are associated with non-abelian statistics. Yet, the spatial extent of Majorana end states measured by S. Nadj-Perge et al. \cite{Yazdani} is restricted to a few atomic distances, making difficult to handle them for braiding. 

An alternative proposal for manipulating Majorana quasiparticles consists in engineering a one-dimensional topological superconductor in a chain of magnetic atoms with a spiral magnetic order on the surface of a superconductor \cite{Nadj-Perge2013,Choy2011,Nakosai2013,Braunecker2013, Klinovaja2013, Vazifeh2013,Pientka2013,Kim2013}. Individual local magnetic moments act destructively on Cooper pairs, leading to discrete spin-polarized states inside the superconducting energy gap, predicted by Yu, Shiba and Rusinov \cite{Yu1965, Shiba1968, Rusinov1969} (YSR). Rusinov suggested that around magnetic atoms the decaying YSR wavefunction should have a spatially oscillating structure \cite{Rusinov1969, Bauriedl1981, Balatsky2006}. The emergence of topological superconductivity depends on the YSR states mediated coupling inside the magnetic chain \cite{Pientka2013}.
While Nadj-Perge et al. \cite{Yazdani} relied on a direct interaction between neighboring atoms to generate short-ranged Majorana quasi-particles, in the latter case the characteristic length is that of the YSR bound states, which may extend up to the scale of the superconducting coherence length. Enhancing the spatial extent of YSR bound states would facilitate the remote coupling of magnetic systems through a superconducting state, opening the route towards an easier manipulation of Majorana quasiparticles and the creation of new topological quantum devices.

\begin{figure*}[ht!] 
\includegraphics[scale = .8]{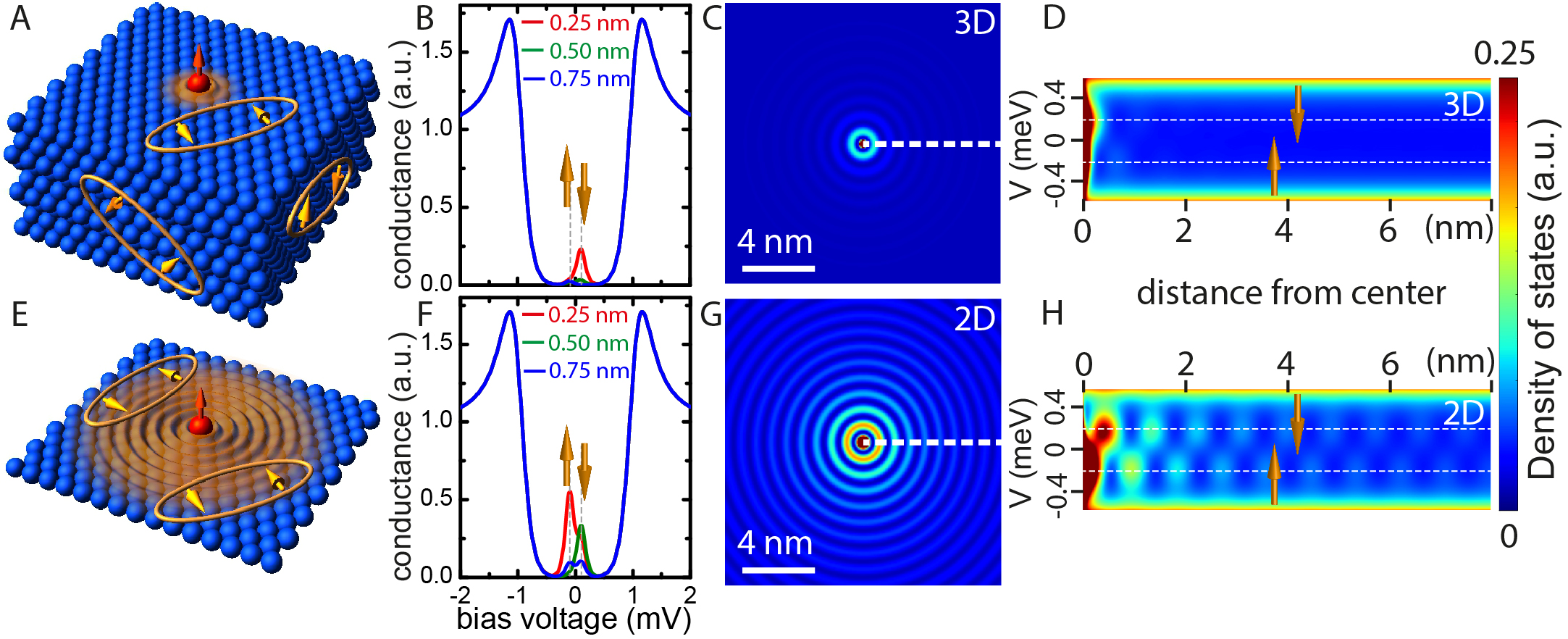}
\caption{\textbf{Comparison between 3D and 2D for the spatial extent of Yu-Shiba-Rusinov states} (A) to (D) ((E) to (H)) Calculated behavior of a Yu-Shiba-Rusinov bound state in an isotropic s-wave superconductor with three-dimensional (two-dimensional) electronic band structure. (A) and (E) are schematic views of the interaction of Cooper pairs with a classical magnetic impurity. (B) and (F) are calculated scanning tunneling spectra at various distances from the impurity showing the fully polarized YSR states inside the superconducting gap. (C) and (G) are simulated conductance maps around the impurity showing the spatial extent of one peak of the YSR state presented in Figs. B. and F. respectively. (D) and (H) are simulated conductance between -0.6 and 0.6 mV along the dotted line out of the impurity in Figs. (C) and (G) respectively.}
\label{Fig1}
\end{figure*}

\begin{figure*}[ht!] 
\includegraphics[scale = .8]{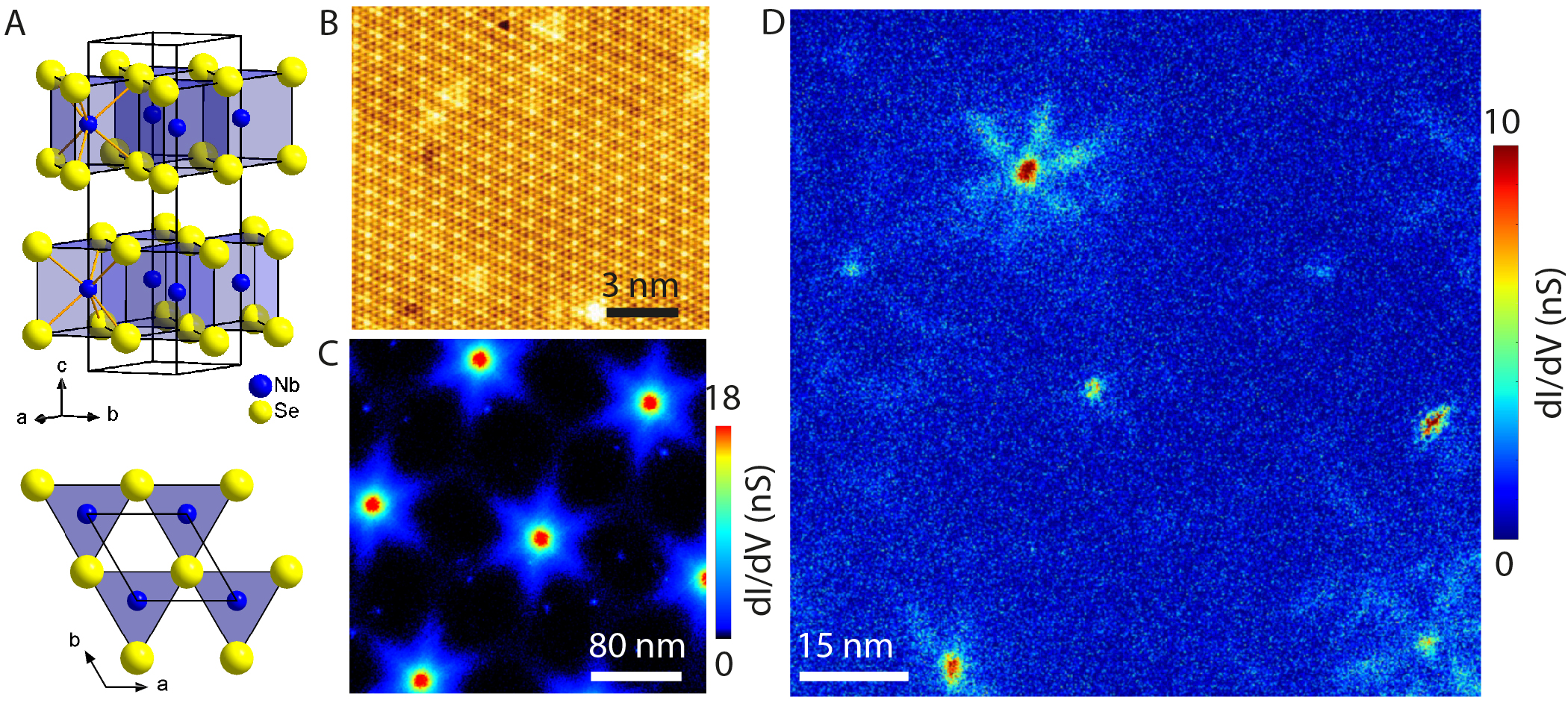}
\caption{\textbf{Structural and superconducting properties of 2H-NbSe$_2$} (A) Atomic structure of a 2H-NbSe$_2$ crystal. (B) 19$\times$17~nm$^2$ topographic image of a 2H-NbSe$_2$ sample with the atomic lattice modulated by a charge density wave. The image is taken at V=-200~meV and I=80~pA. (C) Abrikosov lattice in 2H-NbSe$_{2}$ showing the star shaped structure of vortices in a magnetic field of 0.1~T. (D) Conductance map taken at V=-0.05 meV showing a few star-shaped structures created by localized magnetic impurities at zero magnetic field. Measurements were performed at 320~mK.}
\label{Fig2}
\end{figure*}

Here we reveal that the dimensionality plays a critical role in the spatial decay of the YSR bound states. Calculations using Rusinov's approach \cite{Rusinov1969} are presented in Figs. \ref{Fig1}.A-D. They show that a three-dimensional isotropic s-wave superconductor induces a dramatic decay of the YSR states out of the magnetic impurity. These calculations are in agreement with the atomically short spatial extent of the YSR states observed in all previous scanning tunneling microscopy (STM) studies of single magnetic impurities in superconductors: Co, Cr, Mn, Gd atoms deposited on Pb or Nb crystals \cite{Yazdani1997, Shuai-Hua2008} and manganese-phtalocyanine molecules deposited on Pb \cite{Franke2011}. By extending this theory to two dimensions (2D), we evidence in Figs. \ref{Fig1}.E-G that superconductors with two-dimensional electronic structure should host YSR bound states with spatial extent orders of magnitude larger. Hence layered superconducting materials such as 2H-NbSe$_2$, known for their two-dimensional character, are good candidates for supporting these long range quantum states \cite{Rossnagel2001}. 

In this work, we studied single crystals of 2H-NbSe$_2$ containing a few tens of ppm of magnetic Fe impurities. Atomically resolved topographic STM imaging of the surface (Fig.~\ref{Fig2}.B) exhibits the characteristic charge density wave pattern. Individual impurities appear as bright spots. Whereas on topographic STM images the magnetic and non-magnetic impurities cannot be distinguished, the magnetic impurities are the only ones to present a characteristic spectroscopic signature inside the superconducting gap (see supplementary S1).

The scanning tunneling spectroscopy studies performed at 320~mK, well below the critical temperature of 7.2~K, reveal YSR bound states around the randomly dispersed magnetic iron impurities in 2H-NbSe$_2$ (see supplementary S2). Remarkably, these states are characterized by a six-pointed star shaped electronic signature extending as far as 10 nm from defects, as can be seen in Fig. \ref{Fig2}.D. This is more than 10 times larger than the previously observed extension of a few \AA~for YSR bound states \cite{Yazdani1997, Shuai-Hua2008, Franke2011} and is comparable to the in-plane coherence length of 2H-NbSe$_{2}$. This long range pattern is due to the two-dimensional character of 2H-NbSe$_2$ (see Fig. \ref{Fig1}), compared to the observed short range in three-dimensional materials such as Pb or Nb. This unusually long spatial effect may also be amplified by the fact that the Fe impurities are embedded in the atomic lattice. Therefore they may experience a stronger electronic coupling to the superconducting condensate than the adsorbed impurities used in previous experiments.

\begin{figure*}[ht!] 
\includegraphics[scale = .9]{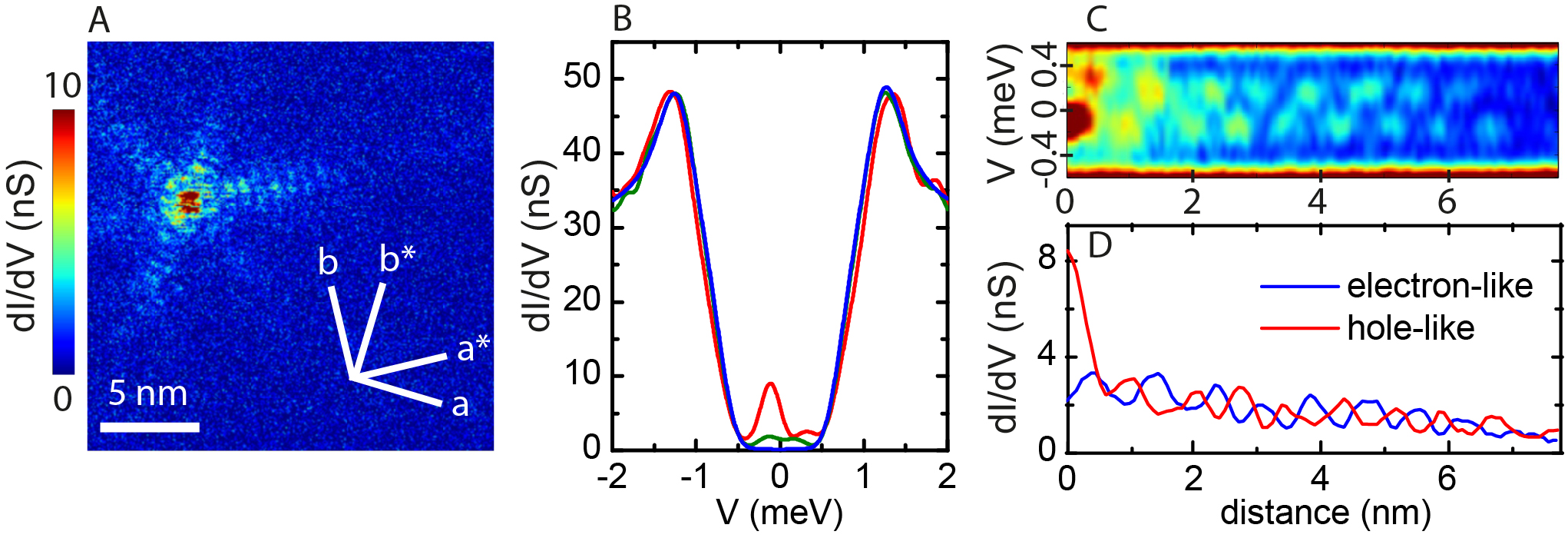}
\caption{\textbf{Spectral and spatial properties of an extended Yu-Shiba-Rusinov bound state in 2H-NbSe$_2$} (A) Experimental conductance map taken at -0.13~meV. Two a and b lines indicate the crystallographic axes of 2H-NbSe$_2$ while the a$^{*}$ and b$^{*}$ indicate the directions in the reciprocal space. (B) Characteristic experimental spectra taken on top of the impurity (red), on the left branch, 4~nm from the center of the impurity (green) and far from the impurity (blue). (C) Spatial and energy evolution of the experimental tunneling conductance spectra, $dI/dV(x,V)$ along one branch of the star. The left side of the figure corresponds to the center of the star and the right side to the top-right corner of the scanning area. The color scale is the same as the one used in (B). (D) Conductance profile of the electron and hole like YSR states as a function of the distance to the impurity along the same line as for (D).} \label{Fig3}
\end{figure*}

The arms of the star shaped pattern of YSR states are turned by 30$^\circ$ with respect to the crystallographic axes of 2H-NbSe$_2$ (Fig. \ref{Fig2}.A), which corresponds to the reciprocal lattice vectors ($\bf{a}^*$ and $ \bf{b}^{*}$). This orientation is also the same as the one of the star-shaped vortices observed in 2H-NbSe$_{2}$ (Fig. \ref{Fig2}.C). In addition to the dominant type of impurities shown on Fig. \ref{Fig2}.D and \ref{Fig3}.A we also observe impurities of Cr and Mn giving star shaped structure with the same orientation but with a thicker pattern and slightly different YSR energies (see supplementary S4).  Therefore this six-fold symmetry is likely to arise from a common origin, and in both cases reflects the anisotropy of the Fermi surface \cite{Rossnagel2001}, as supported by our simulations.

The tunneling spectra acquired over a chosen Fe impurity (see spectroscopic map in Fig.~\ref{Fig3}.A) show a YSR bound state which takes the form of two peaks at positive and negative energies ($E_{Shiba}\simeq\pm$ 0.2 $\Delta$) inside the superconducting gap of 2H-NbSe$_2$ (red curve in Fig.~\ref{Fig3}.B). Apart from the YSR state the characteristic superconducting spectrum is perfectly preserved. The YSR peak at negative bias is much stronger than the one at positive bias, highlighting a strong particle-hole asymmetry near the magnetic atom, as presented in Fig.\ref{Fig1}.H. The presence of a single pair of YSR peaks in the gap indicates that s-wave diffusion channel ($l=0$) dominates, which suggests that the iron impurities may be considered as punctual defects. Similarly, in Gd/Nb, Mn/Nb \cite{Yazdani1997} or Mn-Phtalocyanine on Pb \cite{Franke2011} the $l=0$ diffusion channel was the only one to be activated. On the contrary for Mn/Pb, YSR states for $l=0$ and $l=1$ were observed and $l=2$ states were found for Cr/Pb(111) \cite{Shuai-Hua2008}. However in all these works, the spectroscopic signatures associated with the impurities completely vanished a few \AA~from their center. In this context our measurements show that the local nature of the interaction does not prevent the existence of long range effect on the density of states.

\begin{figure}[b!] 
\includegraphics[scale =1]{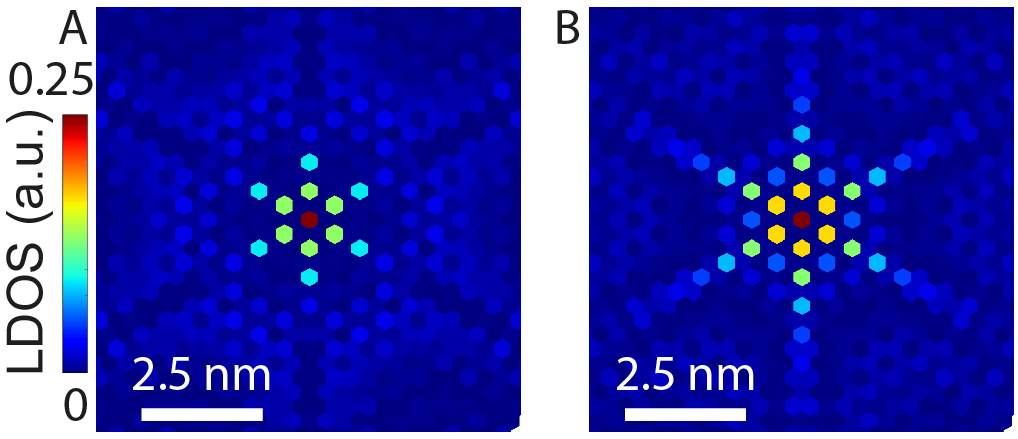}
\caption{\textbf{Tight binding calculation of the spatial structure of Yu-Shiba-Rusinov states} Local density of state (LDOS) computed with  $JS/2=120$ meV and $K=-180$ meV. \textbf{a)} LDOS for electron-like YSR state with energy $E_{Shiba}= 0.59\Delta$. \textbf{b)} LDOS for hole-like YSR state with energy $E_{Shiba}= -0.59\Delta$.}\label{Fig4}
\end{figure}

In order to recover the symmetry of the observed YSR bound state one needs to take into account the band structure of the material. The hexagonal symmetry observed experimentally in Figs. \ref{Fig2}.D and \ref{Fig3}.A is well reproduced in the framework of the Bogoliubov-de Gennes formalism \cite{Flatte2000}. This is done by numerically solving the Schr\"odinger equation with an almost exact tight-binding description of the band structure of 2H-NbSe$_2$ (see supplementary S5). As we only observe $l=0$ states we assume a strictly on-site interaction while treating the magnetic impurity classically, \textit{i.e.} assuming a large spin number $S$ (see supplementary S3). The interaction potential contains both a magnetic and non-magnetic part and reads as
\begin{equation}
H_{Imp}=-\frac{JS}{2}(c_{0\uparrow}^{\dagger}c_{0\uparrow}-c_{0\downarrow}^{\dagger}c_{0\downarrow})+K(c_{0\uparrow}^{\dagger}c_{0\uparrow}+c_{0\downarrow}^{\dagger}c_{0\downarrow})
\label{eq:1}
\end{equation}
Where the $c_0$ and $c_0^{\dagger}$ operators are respectively the annihilation and creation operators for electrons with spin $\sigma$ on the magnetic atom site. The first term corresponds to a Zeeman splitting between spin up and spin down electrons for a coupling strength J/2 between the superconducting electrons and the individual atom. The second term is a non-magnetic diffusion potential of amplitude $K$. Using this approach, we recover the typical star shaped structure as presented on Fig.~\ref{Fig4} aligned along the reciprocal lattice vectors.

We now focus on the details of the star presented on the experimental spectroscopic map, taken at the energy of the strongest YSR peak -0.13~mV
(see Fig.~\ref{Fig3}.A). The high tunneling conductance in the center of the star (red color) corresponds to a very strong peak in the tunneling spectra (red curve on Fig.~\ref{Fig3}.B) localized on the impurity. On the surrounding tail the amplitude of YSR peaks decreases as shown by the green conductance curve of Fig.~\ref{Fig3}.B acquired at 4~nm away from the impurity. This decrease is oscillatory, resulting in interference fringes with a periodicity of 8~\AA~clearly visible on the conductance map. The evolution of the conductance spectra along one arm of the star is shown in Fig. \ref{Fig3}.C. The interference fringes in the conductance for the electron and hole excitations are in an almost perfect spatial antiphase. It appears that a few nanometers away from the impurities the tail of the YSR states exhibits a similar amplitude for the electron-like and hole-like excitations (see Figs. \ref{Fig3}.C-\ref{Fig3}.D), i.e. the particle-hole symmetry is progressively restored away from the defect. Our analysis shows that the oscillations are due to the scattering of electrons by the impurity at the saddle points between the pocket around $\Gamma$ and the pockets around the $K$ points of the Fermi surface of 2H-NbSe$_{2}$.

The observed oscillations being of the order of the Fermi wave-length, they cannot be captured by a discrete tight binding model. Following Rusinov \cite{Rusinov1969} and taking the continuum limit description of the equation giving access to the YSR states, we extract the asymptotic behavior of the wave function far from the magnetic impurity (see supplementary S7). We qualitatively reproduce the two characteristic length scales of the experimentally observed interference pattern. Assuming an isotropic energy band, the YSR energy can be parametrized as  $E=\Delta \cos(\delta^+-\delta^-)$~\cite{Rusinov1969} with $\tan \delta^{\pm}=K\nu_0 \pm JS/2\nu_0$ where $\nu_0$ is the density of states at the Fermi energy. In two dimensions the YSR wave function can be shown to behave at large distance from the impurity as (see supplementary S7):  
\begin{equation}
\psi_\pm(r)=\frac1{\sqrt{N\pi k_F r}} \sin{(k_Fr-\frac{\pi}{4}+\delta^\pm)} e^{-\Delta \sin(\delta^+ - \delta^-) r/\hbar v_F},
\end{equation}
where $\psi_+$ and $\psi_-$ denote respectively the electron and hole components of the YSR wave function $\psi$, $N$ is a normalization factor, $k_F$ is the Fermi wave vector and $v_F$ the Fermi velocity. This behavior presented in Fig.~\ref{Fig1}.H is in excellent agreement with the experimental Figs.~\ref{Fig3}.C-\ref{Fig3}.D. This result highlights the dimensionality dependence as the decay of the local density of states goes as $1/r$ in 2D and $1/r^2$ in 3D (see supplementary S8). Furthermore, for deep YSR states, which corresponds to a dephasing $\delta^+-\delta^-\longrightarrow \pm \pi/2$, the electron and hole YSR states are indeed in antiphase far enough from the impurity. In our approach each component of the YSR states is fully polarized both in spin and charge. The analytical equation links the Friedel-like long distance decay of the YSR state to the superconducting coherence length and the oscillatory behavior to the Fermi wavelength.

%
%


In conclusion, by coupling a classical spin to a superconductor with a two-dimensional electronic structure we were able to unveil a long range coherent magnetic quantum state with a spatially oscillating electron-hole asymmetry. As this effect is related to the dimensionality it should manifest in a wide variety of superconductors, such as lamellar materials or recently discovered superconducting monolayers of Pb/Si(111), In/Si(111) \cite{Zhang2010, Brun2014} and FeSe/SrTiO$_3$\cite{Wang2012}. The interaction between long range YSR states has now to be explored. Then, it could be used for producing new topological phases in hybrid systems. Arrays of magnetic atoms and molecules coupled trough a superconducting medium are indeed expected to present a large variety of topological orders. For instance, a chain of magnetic atoms coupled through the spatially extended YSR bound states with a helical spin order could lead to a topological triplet superconductivity with Majorana quasiparticles at its extremities \cite{Nadj-Perge2013,Choy2011,Nakosai2013,Braunecker2013, Klinovaja2013, Vazifeh2013,Pientka2013,Kim2013}. Such spiral magnetic ordering could be stabilized using the RKKY interaction which oscillates with inter-atomic distance, as for Co/Ir(001) \cite{Menzel2012}. The fine tuning of the YSR and RKKY mediated interactions could be achieved using one- or two-dimensional self-assembly of atoms and molecules on templated substrates \cite{Yazdani, Menzel2012, Fortuna2012, Bazarnik2013}.

{\bf Acknowledgements}
\\
This work was supported by the French Agence Nationale de la Recherche through the contracts ANR Electrovortex and ANR Mistral. G.M. acknowledges funding from the CFM foundation providing his PhD grant. V.S. thanks L.R. Tagirov for his assistance. The authors thank Enric Canadell, Kamran Behnia and Valerii Vinokur for stimulating discussions. 


\end{document}